\documentclass[aoas,MSNbibl,nameyear,rotating,dvips]{arximspdf}
\usepackage{dcolumn}
\usepackage{graphicx}

\doi{10.1214/14-AOAS724} 
\volume{8}
\issue{2}
\pubyear{2014}
\firstpage{824}
\lastpage{851}

\makeatletter
\newcolumntype{d}[1]{D{.}{.}{#1}}

\newcommand{\rrVert}{\Vert}
\newcommand{\rrvert}{\vert}
\newcommand{\llVert}{\Vert}
\newcommand{\llvert}{\vert}
\newtheorem{theorem}{Theorem}

\newtheorem{prop}{Proposition}
\newproclaim{rem}{Remark}
\newproclaim{example}{Example}
\newproclaim{definition}{Definition}
\newproclaim{assumption}{Assumption}
\newproclaim{remark}{Remark}

\newcommand{\margmax}{\mathop{\operatorname{arg\,max}}}

\makeatother

\begin{document}
\begin{frontmatter}

\title{Detection boundary and Higher Criticism approach for rare and
weak genetic effects}
\runtitle{Higher Criticism for rare and weak genetic effects}

\begin{aug}
\author[a]{\fnms{Zheyang} \snm{Wu}\corref{}\thanksref{m1}\ead[label=e1]{zheyangwu@wpi.edu}\ead[label=u2,url]{http://users.wpi.edu/\textasciitilde zheyangwu/}},
\author[a]{\fnms{Yiming} \snm{Sun}\thanksref{m1}\ead[label=e2]{yms@wpi.edu}},
\author[b]{\fnms{Shiquan} \snm{He}\thanksref{m1}\ead[label=e3]{sqhe@wpi.edu}},
\author[c]{\fnms{Judy} \snm{Cho}\thanksref{m2}\ead[label=e4]{judy.cho@yale.edu}},\break
\author[d]{\fnms{Hongyu} \snm{Zhao}\thanksref{m2}\ead[label=e5]{hongyu.zhao@yale.edu}}
\and
\author[e]{\fnms{Jiashun} \snm{Jin}\thanksref{m3}\ead[label=e6]{jiashun@stat.cmu.edu}}

\runauthor{Z. Wu et al.}
\affiliation{Worcester Polytechnic Institute\thanksmark{m1}, Yale
University\thanksmark{m2}\break and Carnegie Mellon University\thanksmark{m3}}
\address[a]{Z. Wu\\
Y. Sun\\
Department of Mathematical Sciences\\
Worcester Polytechnic Institute\\
100 Institute Road\\
Worecester, Massachusetts 01609\\
USA\\
\printead{e1}\\
\phantom{E-mail:\ }\printead*{e2}\\
\printead{u2}}
\address[b]{S. He\\
Worcester Polytechnic Institute\\
100 Institute Road\\
Worecester, Massachusetts 01609\hspace*{7pt}\\
USA\\
\printead{e3}}
\address[c]{J. Cho\\
Internal Medicine\\
Yale University\\
P.O. Box 208056\\
333 Cedar Street\\
New Haven, Connecticut 06520-8056\\
USA\\
\printead{e4}\hspace*{40pt}}
\address[d]{H. Zhao\\
Yale School of Public Health\\
Yale University\\
P.O. Box 208034\\
60 College Street\\
New Haven, Connecticut 06520-8034\\
USA\\
\printead{e5}}
\address[e]{J. Jin\\
Department of Statistics\\
Carnegie Mellon University\\
Baker Hall\\
Carnegie Mellon University\\
Pittsburgh, Pennsylvania 15213\\
USA\\
\printead{e6}}
\end{aug}

\received{\smonth{5} \syear{2013}}
\revised{\smonth{10} \syear{2013}}

%
\begin{abstract}
Genome-wide association studies (GWAS) have identified many genetic
factors underlying complex human traits. However, these factors have
explained only a small fraction of these traits' genetic heritability.
It is argued that many more genetic factors remain undiscovered. These
genetic factors likely are weakly associated at the population level
and sparsely distributed across the genome. In this paper, we adapt the
recent innovations on Tukey's Higher Criticism
(Tukey [The Higher Criticism (1976) Princeton Univ.];
Donoho and Jin [\textit{Ann. Statist.} \textbf{32} (2004) 962--994])
to SNP-set analysis of GWAS, and develop a new theoretical
framework in large-scale inference to assess the joint significance of
such rare and weak effects for a quantitative trait. In the core of our
theory is the so-called \textit{detection boundary}, a curve in the
two-dimensional phase space that quantifies the rarity and strength of
genetic effects. Above the detection boundary, the overall effects of
genetic factors are strong enough for reliable detection. Below the
detection boundary, the genetic factors are simply too rare and too
weak for reliable detection. We show that the HC-type methods are
optimal in that they reliably yield detection once the parameters of
the genetic effects fall above the detection boundary and that many
commonly used SNP-set methods are suboptimal. The superior performance
of the HC-type approach is demonstrated through simulations and the
analysis of a GWAS data set of Crohn's disease.\looseness=-1
\end{abstract}

%
\begin{keyword}
\kwd{Multiple hypotheses testing}
\kwd{large-scale inference}
\kwd{detection boundary}
\kwd{Higher Criticism}
\kwd{rare and weak effects}
\kwd{statistical power}
\kwd{genome-wide association studies}
\kwd{SNP-set methods}
\end{keyword}

\end{frontmatter}

\section{Introduction}\label{sec1}

Genome-wide association studies (GWAS) aim to detect associated genetic
factors by scanning up to several million genetic variants over the
whole genome. Although many genetic factors have been successfully
identified for human diseases, genes discovered to date account for
only a small proportion of overall genetic contribution to many complex
traits [\citet{Kraft2009,McCarthy2008}]. The remaining genetic factors
to be detected likely have weak associations at the population level
and are relatively rare among the huge number of candidates in the
whole genome [\citet{Goldstein2009,Wade2009}]. Besides the efforts to
increase sample size and improve disease classification, it is
desirable to develop statistical methods that more effectively detect
these rare and weak genetic signals not yet discovered.

Two types of statistical association methods are commonly used to
analyze GWAS data: (1) single-SNP methods that analyze the associations
between a trait and individual SNPs, and (2) SNP-set methods that study
the associations between a trait and sets of SNPs. SNP-set methods were
expected to be more promising than single-SNP methods from a biological
perspective. Since multiple SNPs within the same gene, pathway or other
physical and functional genomic segment could jointly affect disease
risk, joint analysis of a set of such SNPs may better reveal the
underlying mechanisms of complex traits than individual SNPs do. In the
past years, many SNP-set methods have been proposed
[\citet{Ballard2010,Hoh2003,Hoh2001,Li2009,LuoXiong2010,Mukhopadhyay2010,Peng2010,Wang2008,Wang2007,Yang2008}]. Despite the encouraging progress in the
literature, there lacks a statistical foundation for when and why the
SNP-set methods would outperform single-SNP methods. In fact, some
SNP-set methods are not automatically better, as we will show in this
paper. At the same time, it is critical to know the limit of any
statistical association methods, as well as the ``best'' of the methods,
especially when genetic effects are rare and weak.

In this paper, we approach these problems from a statistical
perspective. For a set of $L$ SNPs of $n$ individuals, we consider an
additive genetic model
%
\begin{equation}
\mathbf{Y}=\beta_{0}+\beta_{1}\mathbf{X}_{1}+
\beta_{2}\mathbf{X}%
_{2}+\cdots+\beta_{L}
\mathbf{X}_{L}+\bolds{\varepsilon}, \label{equ: genetic model}
\end{equation}
that is frequently used in GWAS [\citet{Kraft2009}]. The linear model is
likely an oversimplification but we develop our
ideas for this one first. See further comments in Section~\ref{Sect:
Discu}. Here $\mathbf{Y}%
=(Y_{1},\ldots,Y_{n})^{\prime}$ is the trait vector, and $\mathbf{X}%
_{j}= ( X_{1j},\ldots,X_{nj} ) ^{\prime}$ is the genotype
vector of
the $j$th SNP, $1\leq j\leq L$.
The error term $\bolds{\varepsilon}= ( \varepsilon
_{1},\ldots,\varepsilon_{n} ) ^{\prime}\sim N ( \mathbf
{0},\sigma
^{2}\mathbf{I} ) $ is independent of the genotypes and can be used
to represent other genetic and environmental variations [\citet
{Falconer1996}]. The variance parameter $\sigma^{2}$ is usually
unknown and needs to be estimated. The coefficient vector $\beta
=(\beta_{1},\beta_{2},\ldots,\beta_{L})^{\prime}$ is unknown to us,
but is presumably rare in the sense that only a few of the coordinates
of $\beta$ are nonzero. We call the $j$th coordinate of $\beta$ a
``signal'' if $\beta_{j}\neq0$ and otherwise a ``noise.'' The term
``rare signal'' should not be confused with ``rare genetic variation.''
Signal rarity is the sparsity among features, but rare genetic
variation is the sparsity among samples. In the literature, while
signal rarity is well defined, signal weakness is a much more vague
notion. As we will show below, signal weakness may result from weak
genetic effect, small sample size and/or small genetic variation.
Signal weakness is one of the main challenges in analyzing big data,
such as GWAS data: The signals are generally very subtle and hard to
find, and it is easy to be fooled.

Statistical literature on linear regression modeling has focused
largely on the goal of separating the signals from the noise [\citet
{Ayers2010,Guan2011,Hoggart2008,WuTT2009,Xie2011}]. While this goal
may provide a perfect solution, it is hard to reach due to a high
demand for strong signals and is often not necessary in GWAS practice
either. Thus, in this paper, we are primarily interested in the problem
of \textit{signal detection}, where the goal is to discover the
associated SNP-sets rather than to identify the individually associated SNPs.

To understand why signal detection is important, from a statistics
point of view it can be shown that given a rarity level of the signals
there is a threshold effect on the signal strength. That is, signals
falling under such a threshold cannot be separated from noise: for any
procedure the sum of the number of signals that are misclassified as
noises and the number of noises that are misclassified as signals
cannot get substantially smaller than the number of signals.
Nonetheless, in many cases while signal rarity and signal strength
prohibit us to separate the signals from the noise, the numerous
\textit
{rare and weak effects} can be combined and utilized in a meaningful
way to solve many challenging problems including, but not limited to,
signal detection, classification and clustering. This challenge has
been successfully met, for example, in \citeauthor{Donoho2004}
(\citeyear{Donoho2004,Donoho2008}),
\citet{Jin2013}. From the genetics point of view, the signal detection problem
is of major interest in the GWAS because the primary target of GWAS is
to screen and allocate the informative genome regions, such as genes,
which are more natural genomic functional units than individual SNPs.
Furthermore, to validate associations, such positive regions will be
further studied and individual SNP effects can still be discovered by
refined and reliable experimental methods.

The signal detection problem in the model (\ref{equ: genetic model})
can be reformulated as a joint hypothesis testing problem where $H_{0}$ is
\[
H_{0}\dvtx\quad \beta_{j}=0,\qquad 1\leq j\leq L,
\]
that is, no association exists between the trait and the SNP sets,
against an alternative hypothesis $H_{1}$ that the trait is associated
with a small
fraction of SNPs in the sets
\[
H_{1}\dvtx\quad \mbox{$\beta_j \neq0$\qquad only for a small
fraction of $j$, $1 \leq j \leq L$}.
\]
See \citet{Donoho2004} for the subtlety of this problem, where the focus
was on a Stein's normal means model, which is much simpler than the
model considered here.

Our study contains two key components: the detection boundary for signal
detection and the statistic of Higher Criticism. We now discuss two
components separately.

The detection boundary can be viewed as a way to address the
fundamental capability and limit of SNP-set methods. In the
two-dimensional phase space calibrating the signal rarity and signal
strength, the detection boundary is a curve that separates the region
of impossibility from the region of possibility. In the region of
impossibility, the signals are so rare and weak that it is impossible
to separate $H_{1}$ from $H_{0}$. That is, even for the most powerful
method available, the signals are so rare and weak that it would have
the sum of types I and~II error rates to be almost $1$. In the
region of possibility, it is possible to separate $H_{1}$ from $H_{0}$,
and there exists a procedure whose sum of types~I and~II error
rates is approximately $0$.

The study of the detection boundary has two merits. First, the
detection boundary is provided as a function of the rarity and strength
of genetic effects, the SNP-set size, the sample size, the error
variance and the allele frequency, and thus simultaneously reveals the
roles of these factors in gene-hunting. The result is applicable to
genetic association studies of both common and rare genetic variants,
the latter are the main target of finding the missing genetic factors
using deep sequencing technologies [\citet{Ansorge2009,Mardis2008,Metzker2010}]. Second, the detection boundary can serve as a benchmark
for evaluating different SNP-set methods. In
particular, note that any procedure will partition the aforementioned
phase spaces into two regions: a region of possibility and a region of
impossibility. We say a method achieves the optimal phase diagram if it
partitions the two-dimensional phase space in exactly the same way as
the optimal procedure does. As a result, for any procedure we can
assess its optimality by investigating whether it achieves the optimal
phase diagram.

Higher Criticism (HC) is a notion that goes back to \citet{Tukey1976},
and it was shown in \citet{Arias2010,Donoho2004},
\citeauthor{Hall2008} (\citeyear{Hall2008,Hall2010}),
\citet{Ingster2010} that HC is useful in detecting very rare and weak effects.
However, these works deal with models different from the genetic model
(\ref{equ: genetic model}), and it is unclear whether the HC continues
to behave well for the
setting considered here. The genetic model is new in several aspects.
First, the covariates are genotype data, rather than standardized or
Gaussian variables. Second, the conditions for correlations among
covariates, that is, the linkage disequilibrium structure, are better
placed on the population correlations, rather than on the empirical
correlations. Third, the error variance is realistically considered as
unknown and needs to be estimated, rather than being assumed as known.

In this paper we adapt the HC to detect rare and weak genetic effects
in a SNP-set analysis context. With substantial efforts, we work out
the exact detection boundary associated with the genetic model (\ref
{equ: genetic model}). We propose a realistic HC procedure for
analyzing real GWAS data and show that it achieves the optimal phase
diagram in a rather broad context. We provide theoretical comparisons
between HC and the several most commonly used SNP-set methods. Somewhat
surprisingly, these well-known SNP-set methods do not achieve the
optimal phase diagram for rare and weak signals. We further demonstrate
the superiority of the HC-type methods with simulated data and real data.

The paper is organized as follows. In Section~\ref{Sect: Genetic Model
and Detection Boundary} we set up the genetic model and provide the
detection boundary for rare and weak genetic effects. In Section~\ref{Sect: HC procedure} an HC procedure is proposed to reach the optimal
detection boundary for rare and weak genetic signals. In Section~\ref{Sect: FDR vs HC} we discuss the connections of HC to False Discovery
Rate (FDR) controlling methods. We show in Section~\ref{Sect: Other
Methods} that some commonly used SNP-set based methods cannot reach the
best detection boundary, and thus are not optimal. In Section~\ref{Sect: Nume} we compare various methods through numerical simulations
and the analysis of a GWAS data of Crohn's disease. In Section~\ref{Sect: Discu} we discuss relevant theoretical and practical issues. The
proofs of the main theoretical results, the fundamental lemmas and
their proofs, as well as the supplementary figures and tables, are
given in the online supplementary material [\citet{wuz2014suppl}].

\section{Genetic model and detection boundary}
\label{Sect: Genetic Model
and Detection Boundary}

In this section we characterize the detection boundary by introducing a
theoretical framework.

We write in model (\ref{equ: genetic model})
\[
\mathbf{X}_{j}=(X_{1j},X_{2j},
\ldots,X_{nj})^{\prime},
\]
so that $X_{kj}$ is the genotype of the $j$th SNP for the $k$th individual,
where $1\leq j\leq L$, $1\leq k\leq n$. Let the minor allele $A_{j}$ of
the $j$th SNP have a minor allele frequency (MAF) of $q_{j}$. We assume
$q_{j}>q>0 $, $1\leq j\leq L$, for some constant $q$. We use the copy number
of minor alleles to code the SNP genotype, which follows a binomial
distribution under the Hardy--Weinberg equilibrium (HWE) [\citet
{Mendel1866,Pearson1904,Yule1902}]
%
\begin{equation}
X_{kj}\sim\operatorname{Binomial} ( 2, q_{j} ).
\label{equ: genotype}
\end{equation}
In some genetic association studies, the individuals are assumed to be
independent, that is, $X_{k_{1}j}$ and $X_{k_{2}l}$ are independent for
any $k_{1}\neq k_{2}$. However, the dependency among SNPs, called
linkage disequilibrium (LD), is a critical feature in GWAS data. We
characterize the LD structure by the correlation matrix $\bolds{\Sigma}=\bolds{\Sigma}_{L\times
L} $ among $X_{k1},\ldots,X_{kL}$. For $\gamma>0$ and $\Delta>0$, let
%
\begin{eqnarray}\label{equ. weak correlation}
&&\mathcal{S}_{L}(\gamma,\Delta)= \{ \bolds{\Sigma}\dvtx \mbox{each row
of } \bolds{\Sigma}\mbox{ has no more than }
\nonumber
\\[-8pt]
\\[-8pt]
\nonumber
&&\hspace*{62pt}\Delta\mbox{ elements exceeding }
\gamma\mbox{ in magnitude} \}.
\end{eqnarray}
With an appropriately small $\gamma$ and a moderately large $\Delta$, a
matrix $\bolds{\Sigma}$ in $\mathcal{S}_{L}(\gamma,\Delta)$ can be
interpreted as
sparse, in the sense that each row of $\bolds{\Sigma}$ has relatively small
coordinates. This setup has been studied in the theoretical statistics
literature [\citet{Arias2010}], and is relatively general and flexible
for GWAS
because the large correlations are allowed between SNPs far from each other.
In Section~\ref{Sect: Other Methods} we will also consider another setup
for $\bolds{\Sigma}$, where the correlation decays polynomially as
the SNP
distance increases.

We develop a theoretical framework where we use $L$ as the driving asymptotic
parameter, and other parameters are tied to $L$ through fixed parameters.
In particular, we model the sample size $n$ by
%
\begin{equation}
n=n_{L}=L^{a}\qquad\mbox{for some constant }a>0. \label{Definen}
\end{equation}
As $L$ grows to $\infty$, $n_{L}$ grows to $\infty$ as well. $n_{L}$ can
be either larger than or smaller than $L$; both cases are common in recent
GWAS.

Next, fixing $1/2<\alpha<1$ which we call the \textit{rarity parameter},
we model the number of associated SNPs by
%
\begin{equation}
K=K_{L}=L^{1-\alpha}, \label{equ. K=L^(1-a)}
\end{equation}
so that the fraction of signals tends to $0$ as $L\rightarrow\infty$. In
our calibrations, $K_{L}\ll\sqrt{L}$ and the signals are very rare.
Seemingly, this is a very subtle situation. In contrast, the case
$0<\alpha<1/2$ is both easier to analyze and less relevant to the major
challenge of the genetic association study, so we omit the discussion
on that. See, for example, \citet{Arias2010,Donoho2004}.

At the same time, let $M^{\ast}\equiv \{ j_{1},\ldots,j_{K} \}
$ be
the support of $\beta$ (or, equivalently, the set of SNPs associated
with $Y$%
), and let $b_{j}$ be the sign of $\beta_{j}$:
\[
b_{j}=b_{j}(\beta)=\operatorname{sgn}(\beta_{j}),\qquad 1
\leq j\leq L,
\]
where $\operatorname{sgn}(x)=0,1,-1$ if $x=0$, $x>0$, and $x<0$,
respectively. From a practical view, the locations and the directions
of the genetic effects are usually unknown, so we assume the
\textquotedblleft worst-case'' scenario and model $b_{j}$ and $M^{\ast
}$ as completely random. In other words, for any fixed indices
$i_{1}<i_{2}<\cdots <i_{K}$, we assume
%
\begin{equation}
P\bigl(M^{\ast}=(i_{1},i_{2},\ldots,i_{K})
\bigr)= \left[\pmatrix{L
\cr
K} \right]^{-1}, \label{DefineM}
\end{equation}
and that given $j\in M^{\ast}$,
%
\begin{equation}
b_{j}=\pm1 \qquad\mbox{with equal probabilities}, \label{Defineb}
\end{equation}
and $b_{j}=0$ if $j\notin M^{\ast}$.

Moreover, let $\tau_{j}$ be the \textit{normalized strength of genetic
effect} at index $j$ by
%
\begin{equation}
\tau_{j}=|\beta_{j}|\sqrt{2nq_{j}(1-q_{j})}/
\sigma, \label{Definetau}
\end{equation}
where we note $\sqrt{2nq_{j}(1-q_{j})}$ is approximately equal to the
$L^{2}$-norm of $\mathbf{X}_{j}$, $1\leq j\leq p$. Together with the
following results, the detection boundary illustrates how sample size
$n$, group size $L$, error deviation $\sigma$, genetic effects $\beta
_j$ and MAF $q_j$ simultaneously determine the detectability of the
genetic signals through a specific function. For example, for rare
variants with reduced $q_j$, the magnitude of their genetic effects
$\beta_j$ need to increase in the same order of $\sqrt{q_j(1-q_j)}$ to
keep the same level of detectability. This result is valuable for
providing a guideline for gene detection in practice.

In the literature [\citet{Arias2010,Donoho2004,Ingster2002}], it is
understood that the most delicate case is for all $j\in M^{\ast}$,
\[
\tau_{j}=O\bigl(\sqrt{2\log(L)}\bigr).
\]
In fact, if $\tau_{j}\gg\sqrt{2\log(L)}$ for all $j\in M^{\ast}$, then
the detection problem is easy and many crude methods can give successful
detection. On the other hand, if $\tau_{j}\ll\sqrt{2\log(L)}$ for all
such $j$, then it is impossible to separate $H_{1}$ from $H_{0}$ and all
methods must fail. In light of this, we recalibrate $\tau_{j}$ through a
so-called \textit{strength parameter} $r_{j}$ by
%
\begin{equation}
\tau_{j}=\sqrt{2r_{j}\log(L)}, \label{Definer}
\end{equation}
where $r_{j}=O(1)$ if $j\in M^{\ast}$ and $r_{j}=0$ otherwise. Write
$\mathbf{r}=(r_{1},r_{2},\ldots,r_{L})^{\prime}$. We have the
following definition.

\begin{definition*}
We call (\ref{Definen})--(\ref{Definer}) the Asymptotic Rare and Weak model
$\operatorname{ARW}(a,\alpha,\mathbf{r})$.
\end{definition*}

The following notation is frequently used in this paper.

\begin{definition*}
A test statistic is said to have \emph{asymptotically full power} if
the sum of its type I and type II error rates converges to 0 for some
critical value. A~test statistic is said to be \emph{asymptotically
powerless} if the sum of its type I and type II error rates converges
to 1 for any critical value.
\end{definition*}

We are now ready to spell out the precise expression of the detection
boundary. The detectability of genetic association between a set of
SNPs and
a trait depends on both the proportion of associated SNPs and the strength
of the genetic effects. The sharp detection boundary (i.e., with the exact
constant) relates the rarity and the strength of the genetic effects
by\vadjust{\goodbreak}
the curve
\[
r=r^{\ast}(\alpha)
\]
in the phase space, where
%
\begin{equation}
r^{\ast} ( \alpha ) =\cases{ %
\alpha-1/2, &\quad
$1/2<\alpha<3/4,$
\vspace*{2pt}\cr
( 1-\sqrt{1-\alpha} ) ^{2}, &\quad $3/4 \leq\alpha<1.$} \label{equ. rho(alpha)}
\end{equation}
The first main conclusion of this paper is that for any fixed $\alpha
\in
(1/2,1)$, if
\[
r_{j}<r^{\ast} ( \alpha )\qquad \mbox{for all $j\in
M^{\ast
}$},
\]
then the genetic effects are merely so rare and weak that it is impossible
to separate $H_{1}$ from $H_{0}$ asymptotically: all statistical tests are
asymptotically powerless!

Later in Section~\ref{Sect: HC procedure}, we show that if there are at
least $L^{-\alpha}$ proportion of genetic effects having
$r_{j}>r^{\ast
} ( \alpha ) $, there exist statistical methods, such as the
HC approach to be discussed, that can reliably detect the genetic
signal with
asymptotically full power.

To rigorously describe our theoretical results, the technique
conditions for asymptotic analysis are summarized as follows. These
assumptions indicate that the SNP correlation matrix $\bolds{\Sigma}$
is sparse and guarantee that $\hat{\bolds{\Sigma}}$ has the same
property as $\bolds{\Sigma}$:

\begin{enumerate}[(A1)]
\item[(A1)] The number of large correlations in each row of $\bolds{\Sigma}
$ is assumed to be $\Delta=O ( L^{\varepsilon} ) $ for all
$%
\varepsilon>0$.

\item[(A2)] The correlation $\gamma$ in (\ref{equ. weak
correlation}) and
the $L$--$n$ relative value $\gamma^{\prime}=\sqrt{\frac{\log L}{n}}$
satisfy some of the following conditions in different theorems for required
levels of sparsity of $\bolds{\Sigma}$:

\begin{enumerate}[(A2.1)]
\item[(A2.1)] $ ( \gamma+\gamma^{\prime} ) L^{1-\alpha
} (
\log L ) ^{4}\rightarrow0$.

\item[(A2.2)] $ ( \gamma^{2}+\gamma^{\prime2} )
L^{1-\alpha
} ( \log L ) ^{3}\rightarrow0$.

\item[(A2.3)] $ ( \gamma+\gamma^{\prime} ) L^{1-\alpha
}\rightarrow0$.

\item[(A2.4)] $\gamma^{3}+\gamma^{\prime3}=O(L^{5\alpha
-4+\varepsilon})$
for all $\varepsilon>0$.

\item[(A2.5)] $\gamma+\gamma^{\prime}=O ( L^{-1/2+\varepsilon
} ) $ for all $\varepsilon>0$.
\end{enumerate}
\end{enumerate}

\begin{theorem}
\label{Thm: all fails} Consider the genetic model setup in (\ref{equ:
genetic model})--(\ref{Definer}). Under assumptions~\textup{(A1)} and \textup{(A2.1)},
all
tests are asymptotically powerless if $r_{j}$ $<$ $r^{\ast} (
\alpha
) $, $j\in M^{\ast}$.
\end{theorem}

%
\begin{figure}

\includegraphics{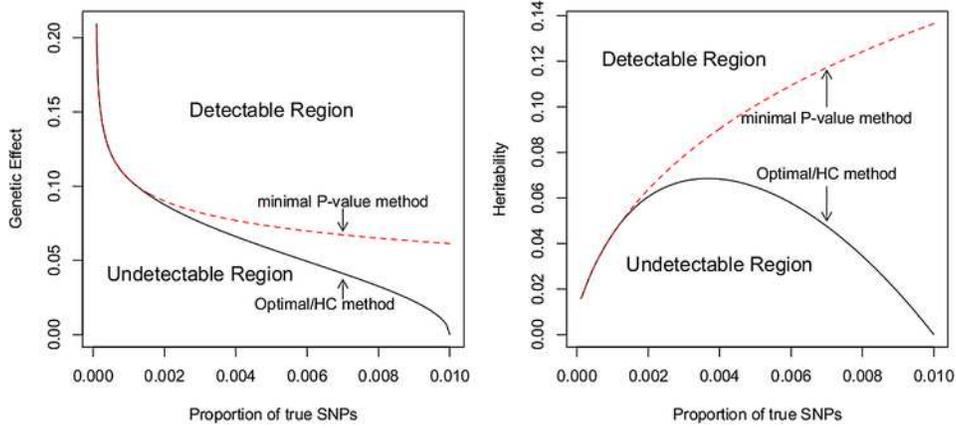}

\caption{Left: Detection boundary on the plane of the proportion of
associated SNPs
and the genetic effect. Right: Detection boundary on the plane of the
proportion of associated SNPs and the heritability. Solid line: the
optimal boundary (reached by HC
procedure); Dashed line: the boundary of the minimal $p$-value method. Here
$L=10\mbox{,}000$, $n=1000$, $\sigma=1$ and $q_{j}=0.3$ for all $j$.}
\label{Figure_DetectionBoundaries}
\end{figure}

By equations (\ref{equ. K=L^(1-a)}) and (\ref{Definetau})--(\ref{Definer}),
for a given proportion of true SNPs $L^{-\alpha}$, the detection
boundary in (\ref{equ. rho(alpha)}) implies the boundaries of detectability
for the genetic effects $\beta_{j}$, as well as for the genetic
heritability of the trait---the proportion of total trait variation due
to genetic variation:
%
\begin{equation}
\mathrm{Heritability}=\frac{\sum_{j=1}^{L}\beta_{j}^{2}2q_{j} (
1-q_{j} ) }{\sum_{j=1}^{L}\beta_{j}^{2}2q_{j} (
1-q_{j} )
+\sigma^{2}}. \label{equ. heritability}
\end{equation}
For easy visualization of these boundaries, consider a special case
where $%
\llvert \beta_{j}\rrvert =\beta$ for $j\in M^{\ast}$ and $%
q_{j}=0.3$ for all $j$. The solid lines in Figure~\ref{Figure_DetectionBoundaries} illustrate
the detection boundary regarding the genetic effect $\beta$ (left
panel) and the detection boundary regarding the heritability (right
penal) over a range of the proportion of associated SNPs corresponding
to $%
\alpha $from $0.999$ to $0.499$.

\section{Higher Criticism procedures for gene detection}
\label{Sect: HC
procedure}

The Higher Criticism (HC) procedure has been studied for the Gaussian
mean model and regression model with Gaussian design matrix and known
error variance [\citet{Arias2010,Donoho2004,Hall2010,Ingster2010}].
Under the genetic model setup in (\ref{equ: genetic model})--(\ref
{Definer}), we adopt this procedure for gene detection based on the
marginal associations between the trait and each SNP. We show that the
HC procedure has asymptotically full power upon the rare and weak
genetic effects exceeding the detection boundary.

Let $p_{ ( 1 ) }\leq\cdots\leq p_{ ( L ) } $ be the
increasingly ordered $p$-values of $L$ individual SNPs. The HC test
statistic is
%
\begin{equation}
\mathrm{HC}_{L}=\max_{1 \leq j \leq L } \mathrm{HC}_{L, j}\qquad
\mbox{where } \mathrm{HC}_{L, j} = \sqrt{L}\frac{(j/L) -p_{ ( j ) }}{\sqrt {p_{ (
j )} (1-p_{ ( j ) }) }}. \label{equ. HC}
\end{equation}

In contrast to considering the minimal $p$-value in a group of SNPs, the
HC considers the maximum of the normalized differences between the
empirical $p$-values $j/L$ and the observed $p$-values $p_{ ( j
) }$.

Denote the survival function of $N(0,1)$ as $\bar{\Phi} ( \cdot
) $. If marginal test statistics $S_j \sim N(0, 1)$, $j=1, \ldots,
L$, and the $p$-values are two-tailed, the HC statistic can be written as
[\citet{Arias2010,Donoho2004}]
%
\begin{equation}
\mathrm{HC}_{L}=\max_{t} \mathrm{HC}_{L}(t)\qquad\mbox{where } \mathrm{HC}_{L}(t) = \frac{|\{j\dvtx | S_{j}| >t \}| - 2L\bar{\Phi}(t)} {\sqrt {2L\bar
{\Phi}(t)(1-2\bar{\Phi}(t))}}. \label{equ. HC equivalent}
\end{equation}

To study the theoretical properties of the HC procedure, for technical
simplification to obtain the upper bound, we follow \citet{Arias2010} to
search for the maximum on a discrete grid and define an HC$^{\ast}$ procedure
with statistic
%
\begin{equation}
\mathrm{HC}^{\ast}_{L}(s) = \max\bigl\{\mathrm{HC}_{L}(t) \dvtx t
\in{}[ s,\sqrt{5\log L}] \cap\mathbb{N} \bigr\}. \label{equ. HC discretized}
\end{equation}
In practice, we recommend to still use the straight HC in (\ref{equ. HC}).

To simplify discussion, we first consider the case where $\sigma^{2}$
is known. For the genetic model in (\ref{equ: genetic model}), let
$\bar{\mathbf{Y}}=  ( \bar{Y},\ldots,\bar{Y} ) ^{\prime}$ and
$\bar{\mathbf{X}}_{j}= ( \bar{X}_{j},\ldots,\bar
{X}_{j}
) ^{\prime}$, where $\bar{Y}=\frac{1}{n}\sum_{k=1}^{n}Y_{k}$ and
$\bar
{X}_{j}=\frac{1}{n} \sum_{k=1}^{n}X_{kj}$. The test statistic $S_j$ for
the association between the trait and SNP $j$ is defined as the
marginal correlation:
%
\begin{equation}
R_{j}^{\sigma}=\frac{ ( \mathbf{X}_{j}-\bar{\mathbf
{X}}_{j} )
^{\prime}\mathbf{Y}}{\sigma\llVert \mathbf{X}_{j}-\bar{\mathbf
{X}}_{j}\rrVert }, \label{equ. R_sigma stat}
\end{equation}
where $\llVert \mathbf{x}\rrVert $ is the $L^{2}$-norm of a
vector $\mathbf{x}$. When SNP $j$ is not associated, we have
$R_{j}^{\sigma} \rightsquigarrow N(0, 1)$.

Proposition~\ref{Thm: varianceknown} states that the HC$^{\ast}$ procedure
reaches the optimal detection boundary. That is, for some
well-controlled type I error rate converging to 0 slowly enough, the
statistical power of the HC$^{\ast}$ procedure converges to 1 for detecting the
genetic effects that fall above the detection boundary.

\begin{prop}
\label{Thm: varianceknown} Consider the genetic model setup in (\ref
{equ: genetic model})--(\ref{Definer}). Let the marginal test statistic
$S_j$ in (\ref{equ. HC equivalent}) be $R_{j}^{\sigma}$. Under
assumptions \textup{(A1)}, \textup{(A2.2)} and \textup{(A2.4)}, $\mathrm{HC}^{\ast}_{L}(\sqrt{2\delta
\log
L})$ with $\delta= \min ( 1,4r^{\ast} ( \alpha )
) $ has asymptotically full power if $r_{j}$ $>$ $r^{\ast} (
\alpha ) $, $j\in M^{\ast}$. Furthermore, under assumptions \textup{(A1)}
and \textup{(A2.5)}, $\mathrm{HC}^{\ast}_{L}(1)$ has asymptotically full power if $r_{j}$
$>$ $r^{\ast} ( \alpha ) $, $j\in M^{\ast}$.
\end{prop}

Now we turn to a more realistic case where $\sigma$ is unknown and
cannot be used in genetic association tests. We propose the following
tests that incorporate $\sigma$ estimation. Specifically, the marginal
association between the trait and SNP $j$ can be measured by either of
the following two test statistics:
%
\begin{equation}
R_{j}=\sqrt{n-1}\rho_{j}\quad\mbox{and}\quad T_{j}=
\sqrt{n-2}\rho_{j}/\sqrt {1-\rho _{j}}, \label{equ. R T stat}
\end{equation}
where $\rho_{j}$ is the Pearson correlation coefficient between the
observed trait values and the genotypes of the $j$th SNP. $T_{j}$ is
the standard $T$-test statistic when we regress the trait on the $j$th
SNP. When SNP $j$ is not associated, both $R_{j}$ and $T_j
\rightsquigarrow N(0, 1)$. Note that $R_{j}$ and $T_{j}$ are asymptotically
equivalent because $\rho_{j}\rightarrow0$ under the $\operatorname{ASW}(a,\alpha
,\mathbf{r})$ model for both the null and the alternative hypotheses.
The numerical results in Section~\ref{Sect: Nume} also show that their
performances are very similar in simulations and real GWAS data analysis.

When $\sigma$ is unknown, we need a slightly stronger condition than
that in Proposition~\ref{Thm: varianceknown} to guarantee the proper
behavior of the $\sigma$ estimation. The following theorem shows that
the HC$^{\ast}$ procedure based on $R_{j}$ still reaches the detection boundary.

\begin{theorem}
\label{Thm: sigmaunknow} Consider the genetic model setup in (\ref{equ:
genetic model})--(\ref{Definer}). Let the mar\-ginal test statistic $S_j$
in (\ref{equ. HC equivalent}) be $R_{j}$. Under assumptions \textup{(A1)},
\textup{(A2.3)} and \textup{(A2.4)}, $\mathrm{HC}^{\ast}_{L}(\sqrt{2\delta\log L})$ with
$\delta=
\min ( 1,4r^{\ast} ( \alpha )  ) $ has
asymptotically full power if $r_{j}$ $>$ $r^{\ast} ( \alpha
) $, $j\in M^{\ast}$. Furthermore, under assumptions \textup{(A1)} and \textup{(A2.5)},
$\mathrm{HC}^{\ast}_{L}(1)$ has asymptotically full power if $r_{j}$ $>$
$r^{\ast} ( \alpha ) $, $j\in M^{\ast}$.
\end{theorem}

Figure~\ref{Figure_DetectionBoundaries} illustrates that the detection
boundary for the HC$^{\ast}$ procedure is the same as the optimal detection boundary.

\section{Connections to FDR-controlling methods}
\label{Sect: FDR vs HC}
Tukey's Higher Criticism (HC) is closely related to methods of
controlling the False Discovery Rate (FDR) [e.g., \citet
{benjamini1995controlling,efron2001empirical}], but is also different
in important ways. While there is a long line of works on FDR
controlling methods, for reasons of space, we focus our discussion on
Benjamini and Hochberg's FDR-controlling method (BH), proposed in \citet
{benjamini1995controlling}. The connection and difference between HC
and BH can be briefly summarized as follows:
\begin{itemize}
\item Both BH and HC are \textit{$p$-value driven methods}, the use of
which needs only the $p$-values associated with all SNPs.
\item BH focuses on the regime where the signals are \textit{rare} but
\textit{relatively strong}, and the goal is signal identification.
\item HC focuses on the regime where the signals are so \textit{rare and
weak} that signal identification is frequently impossible, but valid
signal detection or screening is still possible and could be
substantially helpful.
\end{itemize}
Let $p_{(1)} \leq p_{(2)} \leq\cdots\leq p_{(L)}$ be the sorted
$p$-values associated with $L$ SNPs. The formulas of HC and BH are
intimately connected. In detail, fix the FDR-control parameter $\alpha
\in(0,1)$ (say, $\alpha= 5\%$). The goal of BH is usually to control
the expected fraction of false discovered SNPs out of all discovered
SNPs (i.e., the FDR) so that it does not exceed $\alpha$. The procedure
selects the SNPs whose $p$-values are among the $k_{\alpha
}^{\mathrm{FDR}}$-smallest as discoveries, where $k_{\alpha}^{\mathrm{FDR}}$ is the
largest integer $k$ such that
\[
Q_k \leq\alpha\qquad \mbox{where }Q_k =
\frac{p_{(k)}}{k/L}.
\]
When $\min_{1 \leq k \leq L} \{ Q_k \} > \alpha$, BH reports an empty
set of discoveries.
$Q_k$ is a quantity that has been extensively studied in empirical
processes. See, for example,
\citet{wellner1978limit}.

Following the same argument on page 975 of \citet{Donoho2004}, it can be
shown that for any testing critical value $\sqrt{2 q \log(L)}$ with any
$0 < q < 1$, the ratio between the expected number of recoveries under
the alternative
[with signal slightly above the detection boundary in (\ref{equ.
rho(alpha)})] and the expected number of recoveries under the null is
about 1.
So the problem of BH for the rare and weak signal (which may be
interesting targets in GWAS) is that
%
\begin{equation}
\label{DefineFDR} \min_{1 \leq k \leq L} \{ Q_k \} \approx1.
\end{equation}
As a result, for any $\alpha$ that is bounded away from $1$ (say,
$\alpha\leq90\%$), the BH method reports an empty set of discoveries.
The BH method could produce a nonempty set of discoveries if we let
$\alpha$ get even closer to $1$, but
the FDR is so high that the set of discoveries is no longer informative
for signal identification.

We will never know what was in Tukey's mind when he proposed the Higher
Criticism in 1976 [\citet{Tukey1976}],
but there is an interesting connection between HC and BH (which was
proposed about 20 years later) as follows. Suppose we apply $Q_k$ to
the HC statistic in (\ref{equ. HC}). Heuristically, if $k \ll L$ and
(\ref{DefineFDR}) holds,
\[
\mathrm{HC}_{L, k} \approx\sqrt{k} (1 - Q_k ).
\]
As before, think of the signal detection problem as testing a null
hypothesis $H_0$ versus an alternative hypothesis $H_1^{(L)}$.
In the null case where all $p$-values are i.i.d. from $U(0,1)$ and so
that data contains no signal at all, then $\mathrm{HC}_{L, k} \approx N(0,1)$
for all $k$, and $\mathrm{HC}_{L, k}$ are uniformly bounded from above by a
relatively small number, say, $3$. In the alternative case where the
$p$-values come from rare and weak signals, even when $Q_k \approx1$
for all $k$, it is still possible that for some $k$,
\[
\mathrm{HC}_{L, k} \approx\sqrt{k} (1 - Q_k) \gg1.
\]
This fact says that even when signals are so rare and weak that signal
identification (say, by BH) is impossible, there could still be ample
space for valid inference (e.g., screening or signal detection), and HC
is such a tool. Partially, we guess, this is the reason why Tukey
interprets HC as \textit{the second-level significance testing}.

Denote the maximizing index $k$ for $\mathrm{HC}_{L, k}$ by
\[
k^{\mathrm{HC}} = \margmax_{1 \leq k \leq L} \{ \mathrm{HC}_{L, k} \}.
\]
Such an index is very different from $k_{\alpha}^{\mathrm{FDR}}$. The index
suggests a very interesting phenomenon that is frequently found for
rare and weak signals (however, the phenomenon is not that frequently
found when signals are rare and strong). Specifically, it is not always
the case that $k^{\mathrm{HC}} = 1$; it could happen that the index is larger
than $1$, say, $k^{\mathrm{HC}} = 50$. When this phenomenon happens, the
interpretation is that the strongest evidence against the null is not
necessarily the smallest $p$-value, but is the collection of moderately
smallest $p$-values; see \citet{Donoho2004} for discussion on moderate
significances. When moderate significances contain more information for
inference than does the smallest $p$-value, the HC type methodology is
frequently more appropriate than BH, where the goal is shifted from
signal identification to detection, to accommodate the presence of weak signals.

\section{Some other gene detection procedures}
\label{Sect: Other Methods}

With the genetic detection boundary we can show that many well-known
SNP-set methods are not optimal for the rare and weak genetic effects.
First, we consider the minimal $p$-value method that treats the smallest
$p$-value in a SNP-set as the measurement for the association between the
trait and the SNPs in the set. The following proposition considers the
minimal $p$-value method under cases where $\sigma$ is either known or unknown.

\begin{prop}
\label{Thm: MinP} Consider the genetic model setup in (\ref{equ: genetic
model})--(\ref{Definer}). Under the assumptions \textup{(A1)} and \textup{(A2.2)}, the minimal
$p$-value procedure based on $R_{j}^{\sigma}$ has asymptotically full power
if $r_{j}$ $>$ $r^{\mathrm{MP}} ( \alpha ) $, $j\in M^{\ast}$, and is
asymptotically powerless if $r_{j}$ $<r^{\mathrm{MP}} ( \alpha ) $,
$j\in
M^{\ast}$, where
\[
r^{\mathrm{MP}} ( \alpha ) \equiv ( 1-\sqrt{1-\alpha} ) ^{2},\qquad\alpha
\in ( 1/2,1 ).
\]
Furthermore, under assumptions \textup{(A1)} and \textup{(A2.3)}, the minimal $p$-value
procedure based on $R_{j}$ has asymptotically full power if $r_{j}$ $>$
$%
r^{\mathrm{MP}} ( \alpha ) $, $j\in M^{\ast}$, and is asymptotically
powerless if $r_{j}$ $<r^{\mathrm{MP}} ( \alpha ) $, $j\in M^{\ast}$.
\end{prop}

Proposition~\ref{Thm: MinP} shows that the minimal $p$-value method is not
optimal because $r^{\mathrm{MP}} ( \alpha ) $ $>r^{\ast} (
\alpha
) $ for $\alpha\in(1/2,3/4)$. Figure~\ref{Figure_DetectionBoundaries} illustrates the
comparison between the minimal $p$-value method (dashed curve) and the HC
procedure (solid curve) regarding the genetic effect ($\beta_{j}=\beta$
for all $j\in M^{\ast}$) and the heritability. When the associated
SNPs are
extremely rare with $\alpha\in(3/4,1)$, the two methods have the same
detection boundary. However, in a wide range of the proportion of associated
SNPs corresponding to $\alpha\in(1/2,3/4)$, the HC procedure can detect
significantly weaker genetic effects and heritability than the minimal $p$-value
method does. This regime is more important in combating the detection
of the undiscovered common and rare genetic variants that could number
in the hundreds [\citet{Goldstein2009,Hall2009,Kraft2009,Wade2009}].

We further consider three commonly used SNP-set methods in the GWAS
literature [\citet{LuoXiong2010}] and show that they are not as good as
the minimal $p$-value method under our model setup. Let $\mathbf
{S}=
(S_{1},\ldots, S_{L} ) ^{\prime}$ be a vector of marginal test
statistics and $\hat{\bolds{\Sigma}}$ be the Pearson correlation
coefficients among the SNP
genotypes, that is, $\hat{\bolds{\Sigma}} ( i,j )
=\frac{ ( \mathbf{X}_{i}-\bar{\mathbf{X}}_{i} )
^{\prime
} ( \mathbf{X}_{j}-\bar{\mathbf{X}}_{j} ) }{\llVert
\mathbf
{X}_{i}-\bar{\mathbf{X}}_{i}\rrVert \llVert \mathbf
{X}_{j}-\bar{\mathbf{X}}_{j}\rrVert }$. First, the linear
combination test (LCT) statistic is defined as
%
\begin{equation}
T^{L}=\mathbf{e}^{\prime}\mathbf{S}/\sqrt{\mathbf{e}^{\prime
}
\hat{\bolds{\Sigma}}\mathbf{e}}, \label{equ. TL-test}
\end{equation}
where $\mathbf{e}$ is the vector of 1s. Second, when $\hat{\bolds{\Sigma}}%
^{-1}$ exists, the quadratic test (QT) statistic is defined as
%
\begin{equation}
T^{Q}=\mathbf{S}^{\prime}\hat{\bolds{\Sigma}}^{-1}
\mathbf{S}. \label{equ. TQ-test}
\end{equation}
Third, the decorrelation test (DT) statistic is the Fisher's combination
test after the decorrelation generating independent $p$-values:
%
\begin{equation}
T^{D}=-2\sum_{j=1}^{L}\log
p_{j}, \label{equ. TD-test}
\end{equation}
where the $p$-values $p_{j}=2\bar{\Phi} ( \llvert  W_{j}\rrvert
) $ and $W_{j}$ is the $j$th element of $\mathbf{W}=\mathbf
{D}^{-1}\mathbf{S}$, where $\mathbf{D}$ is a triangular matrix of
Cholesky decomposition such that $\hat{\bolds{\Sigma}}=
\mathbf{D}\mathbf{D}^{\prime} $. The following theorem says that LCT, QT and DT are not
optimal for rare and weak effects when SNPs are independent or have a
polynomially
decaying correlation along the distance between the SNPs. Specifically,
for the true
correlation matrix among the SNPs $\bolds{\Sigma}$, we denote the
operation norm as $\llVert \bolds{\Sigma}\rrVert =\sup_{
\{ \mathbf{a}\dvtx \llVert \mathbf{a}\rrVert _{2}=1 \}
}\llVert \bolds{\Sigma} \mathbf{a}\rrVert _{2}$. $\bolds{\Sigma}$ has a
polynomial off-diagnal decay if for positive constants $M$, $\lambda$
and $C$, the magnitude of the $ ( j,k ) $th element is upper
bounded by a polynomial function
%
\begin{equation}
\bigl\llvert \bolds{\Sigma} ( j,k ) \bigr\rrvert \leq M \bigl( 1+\llvert j-k
\rrvert \bigr) ^{-\lambda}\quad\mbox{and}\quad\llVert \bolds{\Sigma}\rrVert \geq C>0.
\label{equ. poly decay}
\end{equation}

\begin{theorem}
\label{Thm: others less powerful} Consider the genetic model setup in
(\ref{equ: genetic model})--(\ref{equ: genotype}), (\ref
{Definen})--(\ref{Definer}) and (\ref{equ. poly decay}). The three
tests in (\ref{equ. TL-test})--(\ref{equ. TD-test}) correspond to
$\mathbf{S}= ( R_{1},\ldots,R_{L} ) ^{\prime}$, where $R_{j}$ is
defined in (\ref{equ. R T stat}). Let $\gamma^{\prime}=\sqrt{\frac
{\log L}{n}}$. For any $\lambda\geq3$, $M\geq1$ and $\gamma^{\prime
}L=o(1)$, LCT does not have
asymptotically full power when $r_{j}$ $<1$, $j\in M^{\ast}$. For any
$\lambda>1$ and $\gamma^{\prime}L^{d}=o(1)$ for some $d>1$, both QT
and DT do not have asymptotically full power when $r_{j}<1$, $j\in
M^{\ast}$.
\end{theorem}

Because the detection boundary $r^{\mathrm{MP}} ( \alpha ) $ of the minimal
$p$-value method is always less than $1$ for each $\alpha\in(1/2,1)$, the
SNP-set methods LCT, QT and DT have poorer performance than the minimal
$p$-value method. In particular, this theorem indicates that Fisher's
combination test (such as DT) is not a good choice for the rare and weak
genetic effects considered here.

\section{Simulations and Crohn's disease study}
\label{Sect: Nume}

Simulations and real GWAS analysis are conducted to evaluate the
performance of HC-type methods and other traditional and newly proposed
\textit{gene-based SNP-set methods}, in which SNP genotypes in genes
form sets of covariates. Instead of finding individual causative SNPs,
the goal of signal detection here is to test which genes may contain
these causative SNPs. Although the above theoretical results focus on
model (\ref{equ: genetic model}), in order to guide practical
applications, we study both quantitative and binary traits in the
following analysis of three types of data sets (Table~\ref{Table_DataList}): both simulated genotypes and phenotypes, real
genotypes and simulated phenotypes, and both real genotypes and
phenotypes for Crohn's disease study. The following summarizes the
implementation of the methods to be compared:

\begin{longlist}[1.]
\item[1.] Higher Criticism method. The test statistic is given in (\ref
{equ. HC}) for each gene. For quantitative traits, the $p$-values are
calculated based on either $T_{j}$ (method denoted HC) or $R_{j}$
(denoted HCm) in (\ref{equ. R T stat}). For binary traits, we adopt a
$Z$-statistic by \citet{Zuo2006} (denoted HC):
%
\begin{equation}
D_{j} = \sqrt{n}\frac{\hat{p}_{\mathrm{case}}-\hat{p}_{\mathrm{control}}}{\sqrt {2\hat
{p}_{\mathrm{all}} ( 1-\hat{p}_{\mathrm{all}} ) }}, \label{equ. D stat}
\end{equation}
where $\hat{p}_{\mathrm{case}}$, $\hat{p}_{\mathrm{control}}$ and $\hat{p}_{\mathrm{all}}$ are the
estimated MAF in cases, controls and the combined group,
respectively. When the $j$th SNP is not associated, $D_{j}
\rightsquigarrow N (0,1 ) $, the two-tailed $p$-values
$p_{j}=2\bar{\Phi} ( \llvert  D_{j}\rrvert  ) $
are applied to (\ref{equ. HC}) to get the HC statistic.

\item[2.] Minimal $p$-value method (denoted MinP). The association of a SNP
set in a gene is determined by the smallest $p$-value $p_{(1)}$. This is
the most commonly used method in GWAS practice. The $p$-values are
obtained either based on $T_{j}$ in (\ref{equ. R T stat}) for
quantitative traits or $D_{j}$ in (\ref{equ. D stat}) for binary traits.

\item[3.] Principal Component Analysis (PCA) [\citet{Ballard2010,Wang2008}]. To measure the significance of a gene, a $p$-value
is obtained by fitting a multiple regression for quantitative traits
(or a logistic regression for binary traits) by using the least
principal components that count over 85\% variation.

\item[4.] Ridge regression (denoted Ridge) [\citet{He2011}]. SNP covariates
in a gene are fitted with traits by ridge regression at the tuning
parameter that minimizes the prediction error based on cross-validation
(R function lm.ridge). The residual sum of squares describes the
goodness of fit of the model, and thus is treated as the score for the
SNP set. The same procedure is applied to both quantitative and binary
traits for simplicity.\vadjust{\goodbreak}

\item[5.] Linear combination test (LCT), quadratic test (QT) and
decorrelation test (DT) [\citet{LuoXiong2010}]. To calculate the
statistics in (\ref{equ. TL-test})--(\ref{equ. TD-test}), we apply
$\mathbf{S}= ( T_{1},\ldots,T_{L} ) ^{\prime}$ with $T_{j}$ in
(\ref{equ. R T stat}) for quantitative traits and $\mathbf{S}=
(D_{1},\ldots,D_{L} ) ^{\prime}$ with $D_{j}$ in (\ref{equ. D
stat}) for
binary traits.

\item[6.] Kernel-machine test [\citet{Wu2010}]. This is a SNP-set method
that applies the generalized semiparametric models [\citet{Liu2007,Wu2010}] to detect the association of genes. For the additive genetic
model defined in~(\ref{equ: genetic model}), the linear kernel function
is recommended by the authors [\citet{Wu2010}]. So the semiparametric
model is simplified to either a multiple regression model for
quantitative traits (denoted KMT) or logistic regression for binary
traits (denoted LKMT). The genetic association is measured by a
variance-component score statistic [\citet{Zhang2003}]. We apply the R
functions implemented by the authors of this method.
\end{longlist}


\begin{table}[b]
\tabcolsep=0pt
\caption{List of the data used for analysis. Genotypes are either
simulated based on six Toeplitz correlation matrices (TCM) or from the
true GWAS data of NIDDK--IBDGC. The number of SNPs per gene is either
100 or according to the true data. Phenotypes are either simulated
based on the additive model ($\sigma^{2}=1$) or logistic regression
model ($\beta_{0}=-2$) or the true Crohn's disease status. The
locations of nonzero coefficients are always random and the values are
either fixed or random, where $b_{1}$ ranges from 0.088 to 0.131 and
$b_{2}$ ranges from 0.1 to 0.24}\label{Table_DataList}
\begin{tabular*}{\textwidth}{@{\extracolsep{\fill}}lccccccc@{}}
\hline
\textbf{Data} & \textbf{Genotype} & \textbf{Sample} & \textbf
{SNPs/gene} & \textbf{LD} & \textbf{MAF} & \textbf{Phenotype} &
\multicolumn{1}{c@{}}{\textbf{Nonzero coefficients}} \\
\hline
\phantom{0}1 & Simulation & 1000 & 100 & 6 TCM & 0.4 & Additive & 3, $b_{1}$ \\
\phantom{0}2 & Simulation & 2000 & 100 & 6 TCM & 0.4 & Logit & 3, $b_{2}$ \\
\phantom{0}3 & Simulation & 1000 & 100 & 6 TCM & 0.4 & Additive & 3, $+$/$-b_{1}$
equal chance \\
\phantom{0}4 & Simulation & 1000 & 100 & 6 TCM & 0.4 & Additive & 3, $\operatorname{Unif}[b_{1},
1.2b_{1}]$ \\
\phantom{0}5 & Simulation & 1000 & 100 & 6 TCM & 0.4 & Additive & 3,
$\operatorname{Unif}[0.9b_{1}, 1.1b_{1}]$ \\
\phantom{0}6 & \textit{BCHE} & 851 Jew & 100 & real & real & Additive & 3, $b_{1}$
\\
\phantom{0}7 & \textit{BCHE} & 851 Jew & 100 & real & real & Logit & 3, $b_{2}$ \\
\phantom{0}8 & \textit{EXT1} & 851 Jew & 106 & real & real & Additive & 3, $b_{1}$
\\
\phantom{0}9 & \textit{EXT1} & 851 Jew & 106 & real & real & Logit & 3, $b_{2}$ \\
10 & \textit{FSHR} & 851 Jew & 117 & real & real & Additive & 3,
$b_{1}$ \\
11 & \textit{FSHR} & 851 Jew & 117 & real & real & Logit & 3, $b_{2}$
\\
12 & 15,860 genes & 851 Jew & vary & real & real & Additive & $\alpha
=0.8$, $r=0.9$ \\
13 & 15,860 genes & 1145 non-Jewish & vary & real & real & Additive &
$\alpha=0.8$, $r=0.9$ \\
14 & 15,860 genes & 851 Jew & vary & real & real & CD status & -- \\
15 & 15,860 genes & 1145 non-Jewish & vary & real & real & CD status & --
\\
\hline
\end{tabular*}
\end{table}

\subsection{Simulated genotypes and phenotypes}
\label{Sect. SimuSNPdata}

We simulated both genotype and phenotype data to fully control the data
structure and genetic effect pattern. Data sets 1 and 2 in Table~\ref{Table_DataList} were obtained in the following. First, to simulate the
genotype data, it was assumed that one gene unit contains $L=100$ SNPs,
whose genotypes follow HWE in (\ref{equ: genotype}) with MAF $q=0.4$.
To demonstrate how typical LD structures may affect these methods, six
Toeplitz correlation matrices (TCM) were studied: (I) Independent SNPs,
that is, the correlation matrix $\bolds{\Sigma}$ is the identity matrix. (II)
SNPs in the first order neighborhoods are correlated, that is, $\bolds{\Sigma}$
has 1 in the main diagonal, 0.3 (or 0.25, or 0.2) in the first
off-diagonals and 0 elsewhere. (III) SNPs are correlated with the
nearest two neighbors, that is, $\bolds{\Sigma}$ has 1 in the main diagonal,
0.25 in the first off-diagonal, 0.3 (or 0.2) in the second off-diagonal
and 0 elsewhere. The R package mvtBinaryEP [\citet{By2011,Emrich1991}]
was used to generate the correlated
genotype data.

Second, to simulate the phenotype data, we considered the cases of rare
and weak genetic effects based on the above theoretical results.
Specifically, the rarity parameter was assumed $\alpha=0.76$, so
$K=L^{1-\alpha}\approx3$ randomly picked SNPs were made causative.
Quantitative traits were generated by model (\ref{equ: genetic model})
with error variance $\sigma^{2}=1$. The sample size was $n=1000$. We
examined a series of strength parameters $r_{j}=r$ in (\ref{Definer})
from $0.4$ to $0.9$, which correspond to the genetic effects $\beta
_{j}$ in (\ref{Definetau}) equals $b_{1}$ ranging from $0.088$ to
$0.131$, and the heritability of trait ranging in (\ref{equ.
heritability}) from $0.011$ to $0.024$. On the other hand, binary
traits were generated by a
logistic model
%
\begin{equation}
\operatorname{logit} \biggl( \frac{P ( Y=1|\mathbf{X} )
}{P (
Y=0|\mathbf{X} ) } \biggr) =\beta_{0}+
\beta_{1}X_{1}+\beta _{2}X_{2}+\cdots+
\beta_{L}X_{L}.\label{equ: genetic binary model}
\end{equation}
Conditional on the genotype data, many diseased ($Y=1$) and nondiseased
outcomes ($Y=0$) were generated according to the genetic risk. Then the
retrospective case--control data were collected by randomly sampling
$1000$ cases and $1000$ controls. We considered the coefficient $\beta
_{0}=-2$ and a sequence of nonzero coefficients $\beta_{j} = b_{2}$
ranging from $0.10$ to $0.24$, which correspond to the disease allele
odds ratio ranging from $1.11$ to $1.27$.

\begin{figure}[t!]

\includegraphics{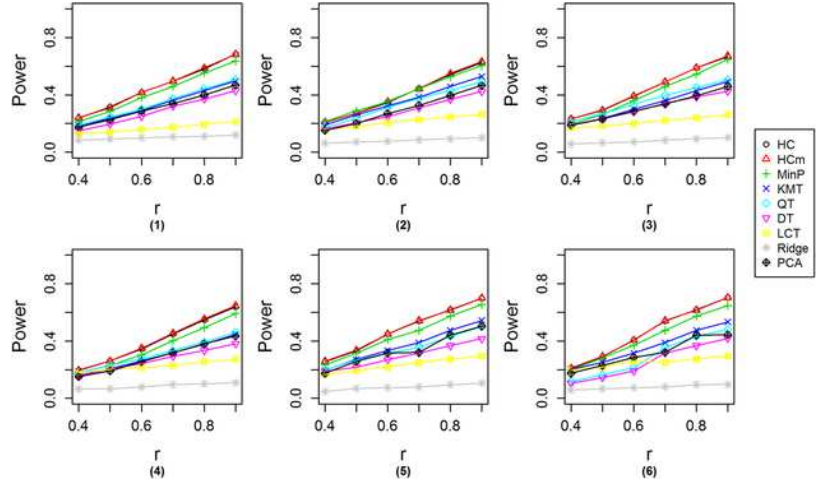}

\caption{For quantitative traits under the fixed value of nonzero
coefficients, HC and HCm have the highest power. X-axis: the strength
parameter $r$ in equation (\protect\ref{Definer}), which corresponds
to the
nonzero coefficients $\beta_{j}=b_{1}$ in (\protect\ref{Definetau}).
The six
panels correspond to six correlation matrices of SNPs: (1) identity
matrix, (2) the 1st off-diagonals equal 0.3, (3) the 1st off-diagonals
equal 0.25, (4) the 1st off-diagonals equal 0.2, (5) the 1st
off-diagonals equal 0.25 and the 2nd off-diagonals equal 0.3, (6) the
1st off-diagonals equal 0.25 and the 2nd off-diagonals equal 0.2.}
\label{Figure_Simu_Quanti}
\end{figure}

\begin{figure}[t!]

\includegraphics{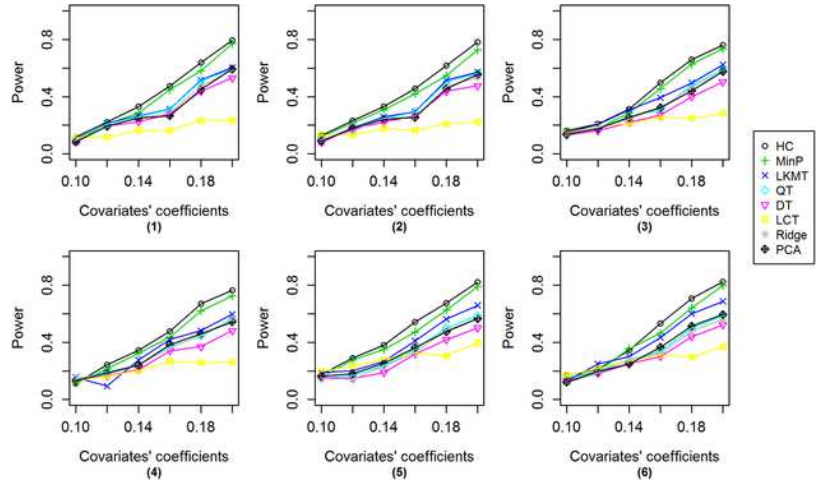}
\caption{For binary traits from the fixed value of nonzero
coefficients, HC has the highest power. X-axis: the nonzero
coefficients $\beta_{j}=b_{2}$ in equation (\protect\ref{equ:
genetic binary
model}). The six panels correspond to the same six correlation matrices
of SNPs as those in Figure \protect\ref{Figure_Simu_Quanti}.}
\label{Figure_Simu_Binary}\vspace*{12pt}
\end{figure}

The empirical power was compared based on a well-controlled empirical
type I error rate. Specifically, we ran 1000 simulations, each with
newly generated genotypes, and then the phenotypes according to a
specific genetic model with random locations of causative SNPs. For
each simulation, we also permuted the phenotype responses and
calculated the test statistics for the null hypothesis of no
association. Over all simulations, the 95th percentile of the null
statistics was used as the cutoffs to control the type I error rate at
a level 0.05. The empirical power, that is, the true positive rate of
tests, is the proportion of simulations where the test statistics
exceeded the corresponding cutoff. Figures~\ref{Figure_Simu_Quanti} and
\ref{Figure_Simu_Binary} show the comparisons of empirical power for
data sets 1 and 2, respectively. In all the setups, the HC-type methods
had the highest power. The comparisons were not significantly affected
by these LD structures.

%

%

\begin{figure}

\includegraphics{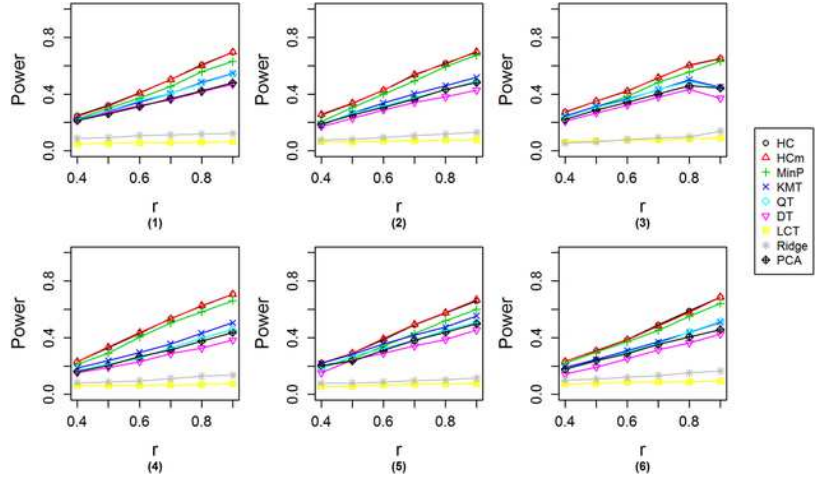}

\caption{For quantitative traits from random nonzero coefficients $\pm
b_{1}$ with equal probabilities, HC and HCm have the highest power.
X-axis: the strength parameter $r$ in equation (\protect\ref
{Definer}), which
corresponds to the nonzero coefficients $\beta_{j}=b_{1}$ in (\protect\ref
{Definetau}). The six panels correspond to the same six correlation
matrices of SNPs as those in Figure \protect\ref{Figure_Simu_Quanti}.}
\label{Figure_Simu_Quanti_RandomSign}
\end{figure}

In reality, causative SNPs may not have homogenous contribution to the
traits. We simulated data sets 3--5 described in Table~\ref{Table_DataList}
for three scenarios of random genetic effects. First,
the nonzero coefficients have the same magnitude $b_{1}$, but with
random $+$/$-$ signs of equal probabilities. Second, the nonzero
coefficients are uniformly distributed in $[b_{1}, 1.2b_{1}]$. Third,
the nonzero coefficients are uniformly distributed in $[0.9b_{1},
1.1b_{1}]$. Figure~\ref{Figure_Simu_Quanti_RandomSign} shows the
comparisons of the methods under random nonzero coefficients $\pm
b_{1}$ with equal probabilities. HC methods were still the best among
these methods assessed. Since the genetic effects have two directions,
the linear combination test (LCT) causes the signals to cancel out and
has low power. The results for the other two scenarios of random
genetic effects (data sets 4--5 in Table~\ref{Table_DataList}) are
given in supplementary Figures~1 and 2 [\citet{wuz2014suppl}].

%

\begin{figure}

\includegraphics{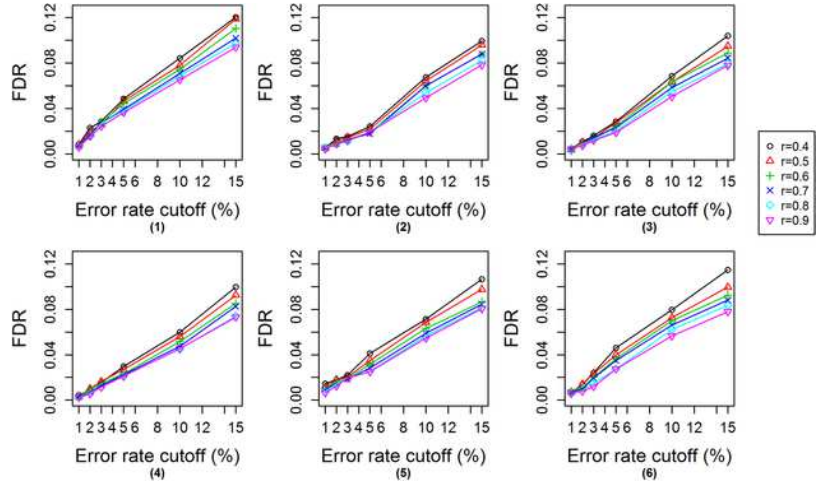}

\caption{False Discovery Rates of the HC method for quantitative
traits. X-axis: the empirical type I error rate cutoff. The six panels
correspond to the same six correlation matrices of SNPs as those in
Figure \protect\ref{Figure_Simu_Quanti}.}
\label{Figure_FDR_HC_Quanti}
\end{figure}

Our theoretical results in Sections~\ref{Sect: HC procedure}--\ref
{Sect: Other Methods} are about reliable detection, that is, to get
asymptotically full power of detecting true genes containing a small
number of weak causative SNPs. In reality, the sample size may not be
large enough to allow the power approaching to 1, and there is a chance
of obtaining false discoveries. Here we assessed the False Discovery
Rate (FDR) of these methods over a variety of type I error rate
cutoffs. Figure~\ref{Figure_FDR_HC_Quanti} illustrates the FDR of HC
methods for quantitative traits (Data 1 in Table~\ref{Table_DataList}),
with the strength parameter $r=0.4$--0.9. It can be seen that the FDR
is well controlled, with an expected decreasing trend for increasing
signal strength $r$. The HC method was also compared with other methods
in terms of the FDR in supplementary Figures~3--8 [\citet
{wuz2014suppl}]. The FDR of the HC method is similar to or lower than
those of the other methods.

%

\subsection{Real genotypes and simulated phenotypes}
\label{Sect. TrueSNPSimuPheno}

By using real genotype data, we studied how the real allelic
distributions and LD structures, which are more complicated than the
above simulations, may influence the results. For this purpose, we used
the observed SNP genotypes from the data of NIDDK--IBDGC (National
Institute of Diabetes, Digestive and Kidney Diseases--Inflammatory
Bowel Disease Genetics Consortium) [\citet{Duerr2006}]. The data contain
851 independent subjects from the Jewish population (417 cases and 434
controls) and 1145 independent subjects from the non-Jewish population
(572 cases and 573 controls). SNPs were grouped into 15,860 genes on
chromosomes 1--22 according to physical locations of genes and SNPs
(NCBI Human Genome Build 35). For data quality control, SNPs were
excluded if they have HWE $p$-values less than 0.01 or MAF less than
0.01. SNPs were also removed if their genotypes are redundant or have a
missing rate over 10\%. The final data set contains 307,964 SNPs. The
gene length (number of SNPs) ranges from 1 to 844 and is highly skewed
to the right: the lower, median and upper quartiles are 3, 7 and 19,
respectively. The missing genotypes were imputed as the average over subjects.

\begin{figure}

\includegraphics{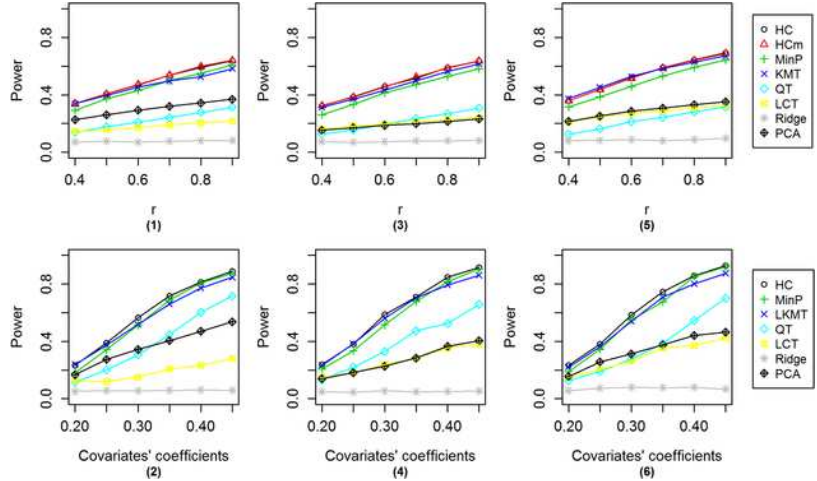}

\caption{Power comparison based on genotype data of genes BCHE (left),
EXT1 (middle) and FSHR (right), respectively. Row 1 X-axis: the
strength parameter $r$ for the genetic effect in equation (\protect\ref
{Definer}) for the quantitative trait model; row 2 X-axis: the genetic
effect $\beta$ in equation (\protect\ref{equ: genetic binary model})
for the
binary trait model.}\label{Figure_Simu_3genes}
\end{figure}

Quantitative and binary traits were simulated under similar setups of
rare and weak genetic effects as those in Section~\ref{Sect.
SimuSNPdata}. Data sets 6--11 in Table~\ref{Table_DataList} list the
parameters and setups based on three genes: \textit{BCHE}
(butyrylcholinesterase) is a gene with 100 SNPs located at
3q26.1-q26.2; \textit{EXT1} (exostosin 1) is a gene with 106 SNPs
located at 8q24.11; \textit{FSHR} (follicle stimulating hormone
receptor) is a gene with 117 SNPs located at 2p21-p16. At the empirical
type I error rate 0.05 from 1000 simulations, Figure~\ref{Figure_Simu_3genes} shows the empirical power of testing these genes
through quantitative (row~1) and binary traits (row 2). It is clear
that HC procedures performed similarly to or better than the other
SNP-set methods.

%

We further studied the performance of these gene-detection methods when
causative SNPs are simultaneously located within multiple risk genes.
Specifically, we took 10 genes found to be associated with Crohn's
disease (CD) in the literature [\citet{Franke2010}] and made each of
these contain $L_{g}^{1-\alpha}$ causative SNPs (rounded to integer),
where $L_{g}$ is the number of SNPs in the $g$th risk gene. The
locations of these associated SNPs in each risk gene were randomly
chosen. The quantitative traits were then generated by an additive
model (\ref{equ: genetic model}) that contains all the causative SNPs
from the 10 risk genes, where each causative SNP has a genetic effect
$\beta_{j}$ defined in (\ref{Definetau})--(\ref{Definer}) with the
rarity parameter $\alpha=0.8$ and the strength parameter $r_{j}=0.9$.
After generating the quantitative trait, we carried out the GWA study
by using the whole genotypes data of all 15,860 genes. Data sets 12 and
13 in Table~\ref{Table_DataList} summarize the information on the
parameters and setups.

To accommodate the fact that genes have distinct numbers of SNPs and LD
structures, we again adapted the permutation test by randomly shuffling
the response traits for obtaining the gene-by-gene empirical $p$-values.
For the 10 risk genes, Tables~\ref{Table_10GeneJew} and \ref
{Table_10GeneNonJew} show their empirical $p$-values from $10\mbox{,}000$
permutations as well as the corresponding ranks (ties are averaged)
among all 15,860 genes based on Jewish and non-Jewish data,
respectively. Only HC methods reliably had the smallest average
$p$-values and ranks for both data sets.

%
\begin{sidewaystable}
\tabcolsep=0pt
\tablewidth=\textheight
\tablewidth=\textwidth
\caption{Based on the NIDDK--IBDGC Jewish genotype data and the additive
genetic model that contains 10 risk genes for Crohn's disease, all
15,860 genes were tested by gene-based SNP-set method, and were ranked
based on their empirical $p$-values. The ranks and $p$-values of the 10
risk genes for each method are listed, and their averages are shown in
the last row}
\label{Table_10GeneJew}
\begin{tabular*}{\textwidth}{@{\extracolsep{\fill}}ld{3.0}d{4.1}d{1.4}d{5.2}d{1.4}d{4.1}cd{4.0}d{1.4}d{4.1}d{1.4}d{4.1}d{1.4}@{}}
\hline
& & \multicolumn{2}{c}{\textbf{MinP}}  & \multicolumn{2}{c}{\textbf{LCT}}  & \multicolumn{2}{c}{\textbf{QT}}
 & \multicolumn{2}{c}{\textbf{KMT}} & \multicolumn{2}{c}{\textbf{HC}}  & \multicolumn{2}{c}{\textbf{HCm}} \\ [-6pt]
& & \multicolumn{2}{c}{\hrulefill}  & \multicolumn{2}{c}{\hrulefill}  & \multicolumn{2}{c}{\hrulefill}
 & \multicolumn{2}{c}{\hrulefill} & \multicolumn{2}{c}{\hrulefill}  & \multicolumn{2}{c@{}}{\hrulefill} \\
 \multicolumn{1}{@{}l}{\textbf{Genes}}&
\multicolumn{1}{c}{\textbf{SNPs/gene}} &
\multicolumn{1}{c}{\textbf{Rank}} &
\multicolumn{1}{c}{\textbf{$\bolds{p}$-value}}&
\multicolumn{1}{c}{\textbf{Rank}} & \multicolumn{1}{c}{\textbf{$\bolds{p}$-value}} &
\multicolumn{1}{c}{\textbf{Rank}} & \multicolumn{1}{c}{\textbf{$\bolds{p}$-value}} &
\multicolumn{1}{c}{\textbf{Rank}} & \multicolumn{1}{c}{\textbf{$\bolds{p}$-value}} &
\multicolumn{1}{c}{\textbf{Rank}} & \multicolumn{1}{c}{\textbf{$\bolds{p}$-value}} &
\multicolumn{1}{c}{\textbf{Rank}} & \multicolumn{1}{c@{}}{\textbf{$\bolds{p}$-value}}\\
\hline
\textit{IL23R} & 23 & 490 & 0.0314 & 1772 & 0.1138 & 2337.5 & 0.1577 &
23 & 0.0007 & 109.5 & 0.0071 & 104.5 & 0.0069 \\
\textit{PTGER4} & 72 & 3984.5 & 0.2496 & 14\mbox{,}246 & 0.901 & 2885 & 0.1931
& 490 & 0.0309 & 470.5 & 0.0313 & 455 & 0.0298 \\
\textit{IL12B} & 41 & 2.5 & 0 & 15\mbox{,}574.5 & 0.9831 & 11.5 & 0.0006 & 3 &
0 & 2.5 & 0 & 2.5 & 0 \\
\textit{CDKAL1} & 160 & 2245.5 & 0.1423 & 4481 & 0.2859 & 6418.5 &
0.4155 & 150 & 0.0084 & 534.5 & 0.0352 & 506.5 & 0.0335 \\
\textit{PRDM1} & 71 & 4801.5 & 0.3029 & 2908 & 0.1858 & 5735.5 & 0.3733
& 8203 & 0.5243 & 8290 & 0.5243 & 8327 & 0.5275 \\
\textit{ZNF365} & 54 & 2.5 & 0 & 1809.5 & 0.1159 & 22 & 0.0013 & 8 &
0.0002 & 2.5 & 0 & 2.5 & 0 \\
\textit{PLCL1} & 64 & 1708.5 & 0.1092 & 8957 & 0.564 & 7353.5 & 0.4751
& 338 & 0.0194 & 807.5 & 0.0505 & 768.5 & 0.049 \\
\textit{BACH2} & 83 & 2.5 & 0 & 9747.5 & 0.6118 & 384 & 0.0274 & 3 & 0
& 2.5 & 0 & 2.5 & 0 \\
\textit{GALC} & 120 & 919 & 0.0578 & 15\mbox{,}391 & 0.972 & 7146.5 & 0.4612 &
1392 & 0.0948 & 936 & 0.0589 & 924.5 & 0.0581 \\
\textit{SMAD3} & 52 & 5806.5 & 0.3642 & 4193 & 0.268 & 3079 & 0.2041 &
5456 & 0.359 & 4985.5 & 0.3135 & 5024 & 0.316 \\[3pt]
Average & 74 & 1996.3 & 0.1257 & 7907.95 & 0.5001 & 3537.3 & 0.2309 &
\multicolumn{1}{c}{\phantom{.}\textbf{1606.6}} & \multicolumn{1}{c}{\textbf{0.1038}} & \multicolumn{1}{c}{\textbf{1614.1}} & \multicolumn{1}{c}{\textbf
{0.1021}} &
\multicolumn{1}{c}{\textbf{1611.9}} & \multicolumn{1}{c}{\textbf{0.1021}} \\
\hline
\end{tabular*}
\end{sidewaystable}

\begin{sidewaystable}
\tabcolsep=0pt
\tablewidth=\textheight
\tablewidth=\textwidth
\caption{Same analysis as that for Table \protect\ref
{Table_10GeneJew}, except
by using the NIDDK--IBDGC non-Jewish genotype data}
\label{Table_10GeneNonJew}
\begin{tabular*}{\textwidth}{@{\extracolsep{\fill}}ld{3.0}d{4.1}d{1.4}d{5.2}d{1.4}d{4.1}cd{4.1}d{1.4}d{4.1}d{1.4}d{4.1}d{1.4}@{}}
\hline
& & \multicolumn{2}{c}{\textbf{MinP}}  & \multicolumn{2}{c}{\textbf{LCT}}  & \multicolumn{2}{c}{\textbf{QT}}
 & \multicolumn{2}{c}{\textbf{KMT}} & \multicolumn{2}{c}{\textbf{HC}}  & \multicolumn{2}{c}{\textbf{HCm}} \\ [-6pt]
& & \multicolumn{2}{c}{\hrulefill}  & \multicolumn{2}{c}{\hrulefill}  & \multicolumn{2}{c}{\hrulefill}
 & \multicolumn{2}{c}{\hrulefill} & \multicolumn{2}{c}{\hrulefill}  & \multicolumn{2}{c@{}}{\hrulefill} \\
 \multicolumn{1}{@{}l}{\textbf{Genes}}&
\multicolumn{1}{c}{\textbf{SNPs/gene}} &
\multicolumn{1}{c}{\textbf{Rank}} &
\multicolumn{1}{c}{\textbf{$\bolds{p}$-value}}&
\multicolumn{1}{c}{\textbf{Rank}} & \multicolumn{1}{c}{\textbf{$\bolds{p}$-value}} &
\multicolumn{1}{c}{\textbf{Rank}} & \multicolumn{1}{c}{\textbf{$\bolds{p}$-value}} &
\multicolumn{1}{c}{\textbf{Rank}} & \multicolumn{1}{c}{\textbf{$\bolds{p}$-value}} &
\multicolumn{1}{c}{\textbf{Rank}} & \multicolumn{1}{c}{\textbf{$\bolds{p}$-value}} &
\multicolumn{1}{c}{\textbf{Rank}} & \multicolumn{1}{c@{}}{\textbf{$\bolds{p}$-value}}\\
\hline
\textit{IL23R} & 23 & 7184 & 0.4638 & 13\mbox{,}952.5 & 0.8833 & 3800.5 &
0.2603 & 4979 & 0.3379 & 5626 & 0.3584 & 5635 & 0.3587 \\
\textit{PTGER4} & 72 & 4627.5 & 0.2965 & 3859 & 0.2509 & 2983 & 0.2048
& 2080 & 0.1396 & 2327.5 & 0.1449 & 2318.5 & 0.1446 \\
\textit{IL12B} & 41 & 35 & 0.0016 & 48 & 0.0026 & 2552.5 & 0.1751 & 167
& 0.0075 & 29 & 0.0014 & 29.5 & 0.0013 \\
\textit{CDKAL1} & 160 & 3 & 0.0001 & 393 & 0.0246 & 3.5 & 0.0001 & 38 &
0.0011 & 4.5 & 0.0002 & 5.5 & 0.0002 \\
\textit{PRDM1} & 71 & 878 & 0.0529 & 9258.5 & 0.5888 & 8300.5 & 0.5427
& 41 & 0.0012 & 543 & 0.0322 & 517.5 & 0.0304 \\
\textit{ZNF365} & 54 & 6080.5 & 0.3912 & 4873 & 0.313 & 4021 & 0.2741 &
7593 & 0.4941 & 6857 & 0.4398 & 6837 & 0.4379 \\
\textit{PLCL1} & 64 & 1071 & 0.0656 & 404.5 & 0.0253 & 9475 & 0.6181 &
1048 & 0.0665 & 768 & 0.0479 & 777 & 0.048 \\
\textit{BACH2} & 83 & 2357.5 & 0.1469 & 11\mbox{,}721.5 & 0.7461 & 1055.5 &
0.069 & 2382 & 0.1591 & 1711 & 0.1065 & 1648.5 & 0.1032 \\
\textit{GALC} & 120 & 379.5 & 0.0232 & 14\mbox{,}902 & 0.9419 & 299.5 & 0.0209
& 45 & 0.0014 & 57.5 & 0.0033 & 58.5 & 0.0033 \\
\textit{SMAD3} & 52 & 119 & 0.0069 & 13\mbox{,}274.5 & 0.8428 & 952 & 0.0632 &
2378 & 0.1588 & 98 & 0.0052 & 96 & 0.0051 \\[3pt]
Average & 74 & 2273.5 & 0.1449 & 7268.7 & 0.4619 & 3344.3 & 0.2228 &
2075.1 & 0.1367 & \multicolumn{1}{c}{\textbf{1802.2}} & \multicolumn{1}{c}{\textbf{0.1140}} & \multicolumn{1}{c}{\textbf{1792.3}} &
\multicolumn{1}{c}{\textbf{0.1133}} \\
\hline
\end{tabular*}
\end{sidewaystable}
%
%

\subsection{Real GWAS of Crohn's disease}

Crohn's disease primarily causes ulcerations of the small and large
intestines, which affects between 400,000 and 600,000 people in North
America alone [\citet{Baumgart2007,Loftus2002}]. To detect novel risk
genes of Crohn's disease, we applied the above gene-based SNP-set
methods to the NIDDK--IBDGC data that contain both real genotypes and
Crohn's disease status as the phenotype (see data sets 14 and 15 in
Table~\ref{Table_DataList}).

The genetic architecture of Crohn's disease remains unclear. One way to
partially compare the above methods for detecting remaining risk genes
is to base on risk genes that have similar properties as those
undiscovered ones. In particular, we studied a set of 41 recently
reported putative genes that likely contain such SNPs with rare and
weak genetic effects to the susceptibility of Crohn's disease [Table~2
of \citet{Franke2010}]. The empirical $p$-values and the corresponding
ranks for these 41 genes are summarized in supplementary Tables~1 and 2
in the supplementary materials [\citet{wuz2014suppl}] for the Jewish
data and the non-Jewish data, respectively. For both data sets the HC
method provided higher average ranks for the 41 risk genes than the
other methods.

For the top 96 ranked genes by HC and those by MinP methods, 87 of them
are common. Nine genes were included in the top 96 genes by HC, but not
by MinP: \textit{PFAAP5}, \textit{AGTR1}, \textit{CDA08}, \textit
{NXPH1}, \textit{LCN10}, \textit{OR51G1}, \textit{FDXR}, \textit
{KIAA1904}, and \textit{EDG1}. Interestingly, by the Catalogue of
Somatic Mutations in Cancer (COSMIC), all nine genes contain one or
more genetic variations associated to the tumor site on the large
intestine. Some of these genes are likely to be relevant according to
their functions. For example, \textit{PFAAP5} (human phosphonoformate
immuno-associated protein~5) on chr13 is likely related to Crohn's
disease, a disease of the immune system. \textit{AGTR1} (Angiotensin II
receptor type 1) on chr3 involves positive regulation of inflammatory
response [\citet{uniprot2012reorganizing}] and is associated with the
increase of immunoglobulin [\citet{wallukat1999patients}]. As a critical
antibody in mucosal immunity, 3--5 grams of immunoglobulin is secreted
daily into the intestinal lumen [\citet{brandtzaeg2004let}]. For
\textit
{NXPH1} (neurexophilin~1) on chr7, neurexophilins are signaling
molecules that resemble neuropeptides by binding to alpha-neurexins and
possibly other receptors. This gene may be relevant because Crohn's
disease can also present with neurological complications. Gene \textit
{LCN10} is potentially relevant because biopsies of the affected colon
of Crohn's patients may show mucosal inflammation, characterized by
focal infiltration of neutrophils, a type of inflammatory cell, into
the epithelium [\citet{Baumgart20121590}]. Gene \textit{EDG1}
(endothelial differentiation gene 1) has regulatory functions in normal
physiology and disease processes, particularly involving the immune,
and influences the delivery of systemic antigens [\citet
{arnon2011grk2}]. Furthermore, genes \textit{AGTR1}, \textit{CDA08},
\textit{OR51G1} and \textit{EDG1} correspond to the components integral
to membranes [\citet{binns2009quickgo}], thus are also linked to Crohn's
disease, which is categorized as a membrane transport protein disorder.
Certainly, further biological validations are needed to confirm how
these genes are related to Crohn's disease.

\section{Discussion}
\label{Sect: Discu}

This paper makes several contributions to the literature. First, it
considers the detection boundary for rare and weak genetic effects in
the GWAS setting. Second, our approach allows for marker dependencies
(LD) and unknown error variance, which are lacking in theoretical
consideration in the literature and are better aligned with practical
GWAS settings. Third, it shows that some of the commonly used SNP-set
methods are suboptimal. Fourth, it proposes a HC-based method to
evaluate the statistical evidence
of association between a set of SNPs and a complex trait. We show that
this method achieves the most power for the specified rare and weak
genetic effect setting. Application of this method to the second wave
of GWAS will likely help researchers identify more trait-associated genes.

Because the values of $R$- or $T$-test statistics in (\ref{equ. R T stat})
depend on the correlations among the genotypic covariates, the HC
procedure for optimal gene detection implicitly incorporates the LD
information into the hypotheses testing. For example, those SNPs
correlated with an associated SNP likely have larger magnitude of their
$R$- or $T$-test statistics and thus smaller marginal $p$-values. So the
maximization procedure in (\ref{equ. HC}) can capture this information
to strengthen the genetic signal. At least in the polynomially decaying
correlations defined in (\ref{equ. poly decay}), this implicit
LD-incorporation is asymptotically more powerful than some commonly
applied procedures that explicitly calculate and incorporate the
correlation matrix into constructing test statistics [\citet
{LuoXiong2010}], as is illustrated by Theorem~\ref{Thm: others less powerful}.

This paper sheds some light on the power of genetic association studies
based on marginal association tests versus joint association tests
[\citet{Genovese2009}]. One interesting discovery of this paper is that
the HC procedure based on marginal association tests has actually
reached the optimal detection boundary for the additive genetic model
in (\ref{equ: genetic
model}). That is, the merit of joint association analysis is probably
not for the additively joint genetic effects, but rather for gene--gene
interactions [\citeauthor{WuZ2009} (\citeyear{WuZ2009,WuZ2011})].

Although we have derived some theoretical results in this paper, and
the general setup may be a reasonable abstraction of the real model,
the assumptions considered are still relatively simple and may not
capture the complexity of the real genetic architecture. For example,
we did not consider potential gene--gene interactions that are believed
to play an important role in biological systems. However, our work does
represent advances over the simpler setup in the literature [\citet
{Arias2010,Donoho2004}],\vadjust{\goodbreak} with the allowance of genotype covariates and
unknown environmental variance. Our theoretical results offer insights
on the relative performance of different methods, which were supported
by results from simulation and practical GWAS.

Our current work can lead to several future research topics in
statistical genetics. The empirical null distribution may depart from
$N(0, 1)$ in large scale data due to unobserved covariates and/or
correlations [\citeauthor{Efron2004} (\citeyear{Efron2004,efron2007correlation,Efron2007})]. It is
important to address how likely this problem could arise in gene-based
detection in GWAS, and how to theoretically and practically address the
issue in detecting sparse heterogeneous mixtures. From a genetics
perspective, first, it would be interesting to study more complex
genetic models, such as those measuring gene--gene interactions. Second,
the proposed HC procedure can be extended to broader applications in
genetic studies. We have illustrated the methods for gene detection
based on SNP-sets grouped within genes. Depending on the scientific
interests, SNPs can also be grouped based on other genomic segments or
based on pathways containing sets of relevant genes [\citet
{LuoXiong2010,Yu2009}]. For example, in a pathway analysis, we can
directly calculate the HC statistics using all individual SNPs within
the pathway. We can also construct a two-level study, in which we
calculate $p$-values for genes, for example, by the goodness-of-fit test
[\citet{Donoho2004}, Section~1.6] for all SNPs within those genes, then
use $p$-values of genes to calculate an HC type statistic for each
pathway. These strategies will be investigated in further research.

\section*{Acknowledgments}
The authors thank the Area Editor, the Associate Editor and the two
referees for many insightful and constructive comments that have
significantly improved the paper. We appreciate the Computing and
Communications Center at Worcester Polytechnic Institute for
computational support.


\begin{supplement}[id=suppA]
\stitle{Supplement to ``Detection boundary and Higher Criticism
approach for rare and weak genetic effect''}
\slink[doi]{10.1214/14-AOAS724SUPP} 
\sdatatype{.pdf}
\sfilename{aoas724\_supp.pdf}
\sdescription{We provide the proofs for main theoretical results, the
fundamental lemmas and their proofs, as well as additional figures and
tables that show performance of Higher Criticism in comparing with
other methods under a variety of setups.}
\end{supplement}

%

\printaddresses

\end{document}